\documentclass{osa-article}
\usepackage{lipsum}

\journal{oe}

\newcommand{\donotshow}[1]{}
\articletype{Research Article}

\begin{document}

\title{Soft thresholding schemes for multiple signal classification algorithm}

\author{Sebastian Acuña,\authormark{1,*} Ida S. Opstad, \authormark{1} Fred Godtliebsen, \authormark{1} Balpreet Singh Ahluwalia,\authormark{1, 2} and Krishna Agarwal\authormark{1}}
\address{\authormark{1}Department of Physics and Technology, UiT The Arctic University of Norway, NO-9037 Tromsø, Norway \\
\authormark{2}Department of Clinical Science, Intervention \& Technology, Karolinska Institute, 17177 Stockholm, Sweden}
\email{\authormark{*}sebastian.acuna@uit.no} 



\begin{abstract}
Multiple signal classification algorithm (MUSICAL) exploits temporal fluctuations in fluorescence intensity to perform super-resolution microscopy by computing the value of a super-resolving indicator function across a fine sample grid. 
A key step in the algorithm is the separation of the measurements into signal and noise subspaces, based on a single user-specified parameter called the threshold.
The resulting image is strongly sensitive to this parameter and the subjectivity arising from multiple practical factors makes it difficult to determine the right rule of selection.
We address this issue by proposing soft thresholding schemes derived from a new generalized framework for indicator function design.
We show that the new schemes significantly alleviate the subjectivity and sensitivity of hard thresholding while retaining the super-resolution ability.
We also evaluate the trade-off between resolution and contrast and the out-of-focus light rejection using the various indicator functions. 
Through this, we create significant new insights into the use and further optimization of MUSICAL for a wide range of practical scenarios. 
\end{abstract}

\section{Introduction}

Conventional optical microscopy is limited in resolution due to diffraction of light. The need to overcome this limit has given rise to super-resolution microscopy techniques, also called optical nanoscopy.
Among these techniques, structured illumination microscopy (SIM) \cite{gustafsson2000surpassing} allows a lateral resolution enhancement by a factor of 2 over the optical resolution limit, stimulated emission depletion (STED) microscopy \cite{hell1994breaking} and single molecule localization (SML) \cite{Betzig1642, rust2006sub, Schnitzbauer2017} can achieve resolutions close to 20 nm, and MINFLUX \cite{Gwosch2020} which combines concepts of SML and STED to achieve even 2 nm resolution. However, these techniques require expensive and complex setups as in SIM and STED, or a high amount of light dose and long acquisition time in SML.

A new family of purely computational nanoscopy techniques has also emerged, where statistical analysis of spatio-temporal fluctuations of fluorescence intensity arising from the photokinetic properties of fluorescent molecules is used for super-resolution. Examples include super-resolution optical fluctuation imaging (SOFI) \cite{dertinger2009fast}, super-resolution radial fluctuations (SRRF) \cite{gustafsson2016fast}, multiple signal classification algorithm (MUSICAL) \cite{agarwal2016multiple}, entropy-based super-resolution imaging (ESI) \cite{doi:10.1021/acsphotonics.5b00307}, 3B \cite{Hu2013}, spatial covariance reconstructive (SCORE) super-resolution fluorescence microscopy \cite{10.1371/journal.pone.0094807}, super-resolution with auto-correlation two-step deconvolution (SACD) \cite{Zhao2018FasterSI} and sparsity-based super-resolution correlation microscopy (SPARCOM) \cite{Solomon:18}.

Most of these methods assume that the emitters are stationary and that the photokinetic properties do not change during the image acquisition. 
MUSICAL, on the contrary, simply exploits the spatio-temporal variations as acquired on a temporal stack of fluorescence microscopy images, irrespective of whether they arise from photokinetics or movement of fluorophores.
It decomposes the stack into a set of vectors called eigenimages (details in section \ref{sec:musical-theory}) and separates them into two orthogonal spaces called the signal and the noise subspaces.
These are used into a special function called the \textit{indicator function} that exploits the fact that the spatial-temporal distribution of fluorophores is encoded in eigenimages through the point spread function (PSF) of the microscope. It is designed to have a high value at an emitter location and lower otherwise, which enables super-resolution when applied  to a grid finer than the original microscopy image.


\begin{figure}[t]
    \centering
    \includegraphics[width=\linewidth]{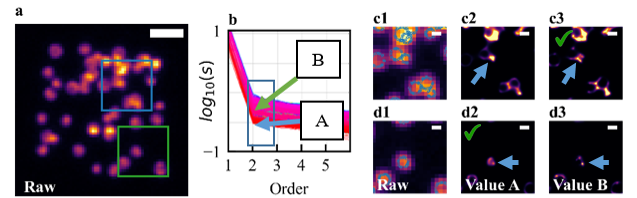}
    \caption{\textbf{Effect of MUSICAL's threshold on image reconstruction.} A sample consisting of rings in the focal plane allows to see the effect of two different thresholds in the final reconstruction (a green tick marks the best one). The sample consists of 500 frames. \textbf{a.} Mean image with 2 regions marked with rectangles. Scale bar 1$\upmu$m. \textbf{b.} Plot of the singular values \cite{Acuna:20} used by the user to pick a threshold. \textbf{c,d.} MUSICAL results in regions 1 and 2, respectively. Scale bars 200 nm. \textbf{c1,d1} show the mean images with the actual emitters on top in blue color.}
    \label{fig:threshold_problem}
\end{figure}

MUSICAL uses three user-specified control parameters that have a bearing on the reconstructed nanoscopy image: a parameter $\alpha$ determines the effective contrast, a sub-pixelation parameter determines the size of the finer grid, and a threshold parameter determines how the eigenimages are split between the signal and noise subspaces.
While $\alpha$ and sub-pixelation prominently determine the aesthetic of the nanoscopy image, the threshold parameter has a significant role in the nature and scale of details that get reconstructed. 
The problem is that its selection depends on multiple factors such as fluorophore's density and photokinetics, sample's geometry, out of focus light, signal to noise ratio, etc. in a complicated and non-obvious manner.
Even though some rules of thumb for threshold selection have been reported \cite{agarwal2016multiple, opstad2019adaptive, Acuna:20}, it still remains as the most non-intuitive parameter of MUSICAL as its value is subjective and depends mostly on the user's experience.
The problem of threshold selection is illustrated in Fig.~\ref{fig:threshold_problem}, where for the same sample (Fig.~\ref{fig:threshold_problem}a), a single value does not produce the best possible reconstruction for the entire sample. 
This figure also illustrates the process of threshold selection, which starts with a plot similar to Fig.~\ref{fig:threshold_problem}b from where the user selects a value such as -0.5 (point A) or -0.3 (point B).
The name of the points come from the rules used to pick them (more details in section \ref{sec:thresh_ori}) and as it can be seen from the results on region 1(Fig.~\ref{fig:threshold_problem}c) and 2 (Fig.~\ref{fig:threshold_problem}d), the results vary between regions.
For example, if we consider the indicated ring in each case, for region 1 the ring is only recovered for value B, while is reconstructed better for value A in region 2 as the ring appears clearer.

This work addresses the problem of the subjectivity of threshold selection and the  sensitivity of MUSICAL's reconstructions to it.
This is achieved in the following manner:

\begin{itemize}
\itemsep0em 
    \item We have scrutinized the effect of threshold selection and identified the root cause of threshold sensitivity.
    \item We have proposed a generalized form of indicator function design that carry two new families of indicator functions.
    \item We have evaluated quantitatively and qualitatively the comparative advantages of the new indicator functions while generating new insights into practically useful properties such as resolution-contrast trade-offs, out-of-focus light rejection, and dynamic range utilization. 
\end{itemize}

The outline of the paper is as follows. 
Section~\ref{sec:background} presents the theory behind MUSICAL and the role of the threshold in image reconstruction.
The new generalized framework for the MUSICAL indicator function design is presented in
Section~\ref{sec:new-method}.
Section \ref{sec:results} presents the results and insights on a variety of simulated and experimental data.
The conclusions are summarized in Section~\ref{sec:conclusion}.

\section{Background} \label{sec:background}
\subsection{The imaging model}
Let's consider a sample composed of point-like blinking photon emitters, i.e. individual fluorophore molecules.
In this case, blinking means that the number of photons emitted by each particle varies over time, owing to the photokinetics of the fluorophores \cite{dempsey2011evaluation}, with no assumption on temporal sparsity. A stack of images taken of such sample from a diffraction-limited system over $T$ time-steps is expressed in a matrix form as $\mathbf{A} = [\mathbf{a}(1) \dots \mathbf{a}(T)]$, where each column vector $\mathbf{a}(t)$ contains the intensity measured by a set of sensing elements (i.e. camera pixels) at time step $t$. For $N$ emitters, a single image $\mathbf{a}(t)$ can be modelled as follows:
\begin{equation}\label{eq:im-model}
    \mathbf{a}(t) = \sum_{n=1}^N \mathbf{g}(\mathbf{r}_n(t))s_n(t).
\end{equation}
In this model, $\mathbf{g}(\mathbf{r})$ represents the PSF of the microscope and corresponds to the intensity produced by an emitter located at $\mathbf{r}$ as measured on the sensor. 
Finally, $s_n(t)$ is the number of photons produced by the $n^{th}$ emitter during the time step $t$, which is a random variable resulting from the photokinetics of the fluorophore.
For simplicity, we here only consider stationary emitters, $r_n(t)=r_n$. Then, Eq.~(\ref{eq:im-model}) can be written as the matrix equation $\mathbf{a}(t) = \mathbf{G}\mathbf{s}(t)$ by constructing two new matrices $\mathbf{G} =[\mathbf{g}(\mathbf{r}_1) \dots \mathbf{g}(\mathbf{r}_N)] $ and $\mathbf{s}(t) = [s_1(t) \dots s_N(t)]^T$. Therefore, every single image is a linear combination of the columns of $\mathbf{G}$, weighted by the photon emissions of emitters.

\subsection{The key concept of MUSICAL} \label{sec:musical-theory}
MUSICAL involves a sliding window operation, with window defined as a crop of the image stack.
Processing a single window returns a super-resolved version of it, and the final MUSICAL reconstruction is built by overlaying and stitching all super-resolved windows together.
The size of the window corresponds to the approximate size of the main lobe of the PSF, which is estimated from the wavelength of emission and numerical aperture of the imaging system. 
While performing nanoscopy using MUSICAL on a single window, the observed data is decomposed into two orthogonal subspaces.
In this paper, we use notation associated with singular value decomposition such as used in \cite{agarwal2016multiple}. 
Accordingly, the matrix $\mathbf A$ is decomposed as $\mathbf A = \mathbf U \mathbf S \mathbf V^T$, where $\mathbf U$ contains the basis vectors ${\mathbf u}_i$. 
These vectors are called eigenimages and contain the spatial information of the sample, while their corresponding singular values $\sigma_i$ in the diagonal of matrix $\mathbf S$ is a measure of its statistical significance.
Since matrix $\mathbf V$ is not used, MUSICAL can benefit of the relation between SVD and Eigenvalue decomposition as $\mathbf A \mathbf A^T = \mathbf U \mathbf \Lambda \mathbf U^T$ (as shown in \cite{Acuna:Thesis:2019} for computational efficiency), where $\Lambda$ contains the eigenvalues $\lambda_i$ instead of the singular values.
Since both values are related ($\lambda_i=\sigma_i^2$), the discussion below can be generally applicable irrespective of whether singular value or eigenvalue decomposition is used. The two key concepts of MUSICAL are presented below.

\vspace{1mm}
\noindent\textbf{Key concept 1 (KC1):} MUSICAL separates the basis $\mathbf U$ into two subspaces, namely the signal subspace $\mathcal{S}$ and the orthogonal complementary noise subspace $\mathcal{N}$ using a threshold $\sigma_0$ such that:  
\begin{equation}
    \mathbf{u}_i \in \begin{cases}
        \mathcal{S} & {\rm{if}}\,\, \sigma_i \geq \sigma_0 \\ 
        \mathcal{N} & {\rm{otherwise.}} 
      \end{cases}
      \label{eq:thresh}
\end{equation}

\noindent\textbf{Key concept 2 (KC2):} If the threshold is chosen such that the subspace $\mathcal{N}$ indeed contains contribution from only the noise, then the subspace $\mathbf{S}$ contains $\mathbf{g}(\mathbf{r}_n)$ for every emitter and $\mathcal{N}$ is devoid of them. Consequently, using the orthogonality of $\mathcal{S}$ and $\mathcal{N}$, $\mathbf{g}(\mathbf{r}_n)$ are also orthogonal to $\mathcal{N}$. This property is used to design the indicator function $f(\mathbf{r})$, discussed in detail in section \ref{sec:indicator_ori}. 
\vspace{1mm}

In the ideal case, the matrix $\mathbf A$ is rank deficient (i.e., it contains some zero singular values), which happens when the number of emitters is smaller than the number of sensor elements. 
In such case, the threshold is simply $\sigma_0 = 0$ and the signal subspace $\mathcal{S}$ is formed by the eigenimages with non-zero singular values while the noise subspace $\mathcal{N}$ by the ones with value zero.
However, real samples are composed of a large number of emitters such that the matrix $\mathbf A$ is full ranked. In addition, real microscopy data contains noise, such that none of the singular values are strictly zero. Therefore, $\sigma_0$ is chosen on a case-by-case basis. Intuitively, the eigenvectors associated with large eigenvalues represent the more statistically prominent structures in the stack as compared to the other eigenvectors.

\donotshow{\textbf{The key concept 2}: Since the space spanned by $\mathbf{g}(\mathbf{r}_i)$ and the eigenimages of the signal space are the same, the noise space is also orthogonal to $\mathbf{g}(\mathbf{r}_i)$, i.e. when the PSF is evaluated at an emitter's location. This property is exploited in designing indicator function that indicates the presence of emitters.} 

\subsection{Indicator function}\label{sec:indicator_ori}
Let us consider an arbitrary test point $\mathbf{r}_{test}$ in the sample region. Its image is given by the vector $\mathbf{g}(\mathbf{r})$, which can be represented as $\mathbf{g}(\mathbf{r}) = \mathbf{g}_\mathcal{S}(\mathbf{r}) + \mathbf{g}_\mathcal{N}(\mathbf{r})$, where $\mathbf{g}_\mathcal{S}(\mathbf{r})$ and $\mathbf{g}_\mathcal{N}(\mathbf{r})$ are the projections of $\mathbf{g}(\mathbf{r})$ onto the signal and noise spaces. The magnitudes of these projections are computed as: 
\begin{equation} \label{eq:dpr}
    ||\mathbf{g}_\mathcal{S}(\mathbf{r}_{test})|| =\sqrt{ {\sum_{\mathbf{u}_i \in \mathcal{S}}{g_i}^2}} 
    \quad ;  \quad 
    ||\mathbf{g}_\mathcal{N}(\mathbf{r}_{test})|| = \sqrt{{\sum_{\mathbf{u}_i \in \mathcal{N}}{g_i}^2}} \quad {\rm where} \quad g_i=|\mathbf{g}(\mathbf{r}_{test})\cdot \mathbf{u}_i|
\end{equation}
Using KC2 and the projections in Eq.~(\ref{eq:dpr}), MUSICAL constructs the following indicator function: 
\begin{equation} \label{eq:indicator-function}
    f(\mathbf{r}_{test}) = \left ( \frac{||\mathbf{g}_\mathcal{S}(\mathbf{r}_{test})||}{||\mathbf{g}_\mathcal{N}(\mathbf{r}_{test})||}\right) ^\alpha
\end{equation}
This indicator function generates a large value when $\mathbf{r}_{test}$ is at the location of the emitters, since $||\mathbf{g}_\mathcal{N}(\mathbf{r}_{test})||$ is zero (KC2). For this it is important that the threshold value $\sigma_0$ for defining the signal and noise subspaces is chosen appropriately.

\begin{figure}[t]
    \centering
    \includegraphics[width=\linewidth]{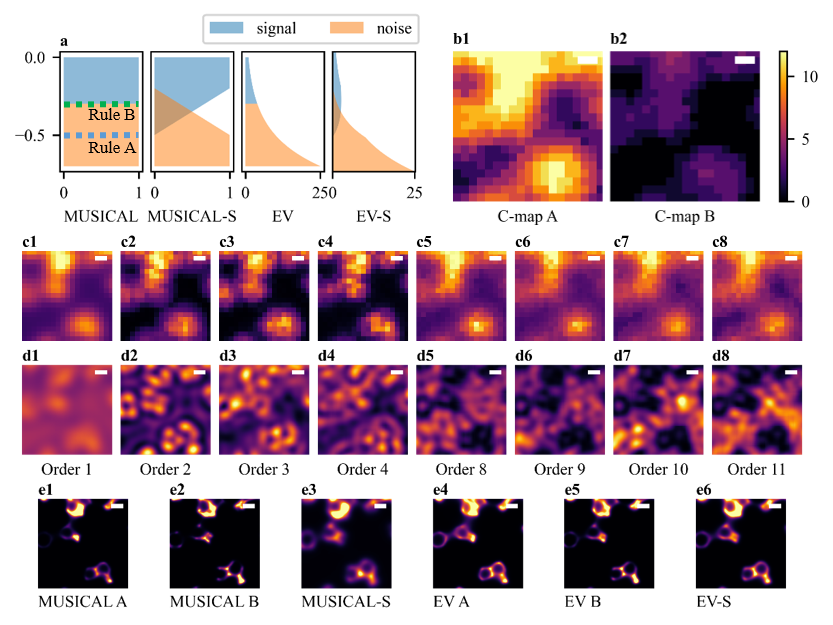}
    \vspace{-8mm}
    \caption{\textbf{MUSICAL's singular values and eigenimages.} \textbf{a.} Weighting coefficients for the different methods. The rules are labeled on top to show how the weights change depending on the threshold. In this example, rule A means a threshold of $-$0.5 while rule B means $-$0.3. For soft methods, the noise and signal are used both in numerator and denominator of the indicator function.  \textbf{b.} Cardinality maps that indicate the number of eigenimages picked as signal for the two rules of selection for every windows (each pixel corresponds to one window). \textbf{c, d.} Heatmap of singular values (c) and projection of different eigenimages (d) sorted by increasing order. \textbf{e.} Results of using different indicator functions. The scale bars are 200 nm.}
    \label{fig:threshold_maps}
\end{figure}

\subsection{Threshold selection and its associated problem}\label{sec:thresh_ori}
The current practice of threshold selection starts with a plot of the singular values of the microscopy image stack, such as shown in Fig.~\ref{fig:threshold_problem}b.
This plot is made by computing the singular values for each window and then plotting them as lines in logarithmic scale, with the x-axis showing the order when sorted decreasingly.
Then, the user specifies a threshold $\sigma_0$ based on observations derived from this plot with particular interest in the inflection or knee point (indicated with a blue rectangle) observed at the second singular value of every window. 
The rule of thumb reported in the original MUSICAL article \cite{agarwal2016multiple}, referred to as Rule A here, involves choosing roughly the threshold as the smallest second singular value across all the windows, i.e. 
\begin{equation}\label{eq:ruleA}
   \textrm{Rule A: } \sigma_0= \min (\sigma_{2}; \forall \rm{windows}).
\end{equation}
According to another rule reported in \cite{opstad2019adaptive}, referred here as Rule B, the threshold is roughly selected as the center of the span of the second singular values for all the windows, i.e.
\begin{equation}\label{eq:ruleB}
   \textrm{Rule B: } \sigma_0= \frac{\min (\sigma_{2}; \forall \rm{windows}) + \max (\sigma_{2}; \forall \rm{windows})}{2}.
\end{equation}

Both rules assume that the eigenimages can be clearly separated and therefore, they constitute hard thresholding schemes.
An illustrative example is shown in Fig.~\ref{fig:threshold_maps}a where the classification and corresponding weighing is shown for Rule B in comparison to Rule A.
In more practical terms, a higher value of threshold, decreases the cardinality of the signal subspace across the entire images as shown in the cardinality maps (c-map) of Fig~\ref{fig:threshold_maps}b.
These maps encode the number of eigenimages in the signal subspace for each window as an intensity value, which is reason why the the c-map B (corresponding to Rule B) looks dimmer.
For even further insight, we show in 
Fig~\ref{fig:threshold_maps}c and d, the actual singular values and eigenimages projection respectively for several orders.
Something interesting to note, is that foreground regions generally have higher eigenvalues than the background for any given order.
However, the projections of $g_i$ shown Fig~\ref{fig:threshold_maps}d follow two different trends. 
First, the value of $g_i$ observe an inversion in the pattern as the order increases. 
The lower orders have high values of $g_i$ in the foreground relative to the background, and vice versa for the higher orders. 
This reversal of trend for higher orders (low singular values) gets exploited in the indicator function as their contribution to the denominator is small. 
Second, generally speaking, eigenimages with lower eigenvalues (i.e. higher orders) are associated with higher spatial frequency components \cite{agarwal2017eigen}, an effect that is more prominent for the first few orders. These observations will be further discussed later in relation with the new indicator functions.

Since Rule B decreases the cardinality of the signal space, it also discards noise much more effectively. 
However, information related to the actual structure may also be relegated as noise, which increases the value of denominator in the foreground and compromises resolution. 
On the other hand, and following the example of Fig~\ref{fig:threshold_maps}, Rule A includes eigenimages up to order 11 in the foreground. 
Yet, as noted in Fig. \ref{fig:threshold_maps}d, the $g_i$ of the foreground is smaller than the background in orders 10 and 11. 
Therefore, their exclusion from the noise subspace is not useful for the foreground. 
Even more: the inclusion of higher $g_i$ corresponding to the background due to these orders may increase the background artifacts.
Therefore, a trade-off is involved in either rule and the manifestation of this trade-off varies from case-to-case. 
 
In practice, even the most experienced bioimaging user of MUSICAL may not know what to expect from the sample being imaged. Moreover, one could argue that more
candidate rules can be designed from analysis of the histogram of the second singular values, such as explored in \cite{Acuna:Thesis:2019}. However, in all these cases, a fundamental limitation is that they all imply hard thresholding: eigenimages included in the signal subspace are considered in a hard solely as representing the structure and the eigenimages included in the noise subspace are considered in solely as representing the noise. 
In reality, the presence of noise implies that each eigenimage is corrupted. 
Therefore, a perfect separation of the eigenimages into a signal or noise subspace is not possible. 

\section{New indicator function design} \label{sec:new-method}

Here, we consider two solutions to the problem mentioned previously:
\begin{itemize}
\itemsep0em 
    \item Eigenvalue (EV) weighing: the magnitude of the eigenvalue is included in the indicator function. It keeps the hard thresholding but softens the effect of first few eigenimages that are classified as noise but may have structural information, as shown in Fig. \ref{fig:threshold_maps}a. 
    \item Soft-thresholding: hard thresholding is removed using a weighing function for each eigenimage. It can be added to MUSICAL and EV, with the new methods abbreviated MUSICAL-S and EV-S, respectively.
\end{itemize}

We begin with a generalized form of indicator function, which allows for more flexibility in its design and that paves the path for developing the indicator functions for the above identified solutions. 
%
%
The most general form of indicator function is shown below:
\begin{equation} \label{eq:indicator-function-general}
    f(\mathbf{r}_{test}) =\left(\sqrt{\frac{\sum_{i=1}^{N}{a_i g_i^2}}{\sum_{i=1}^{N}{b_i g_i^2}}}\right)^
    \alpha
\end{equation}
Here, $a_i$ and $b_i$ are the design parameters of the indicator function.
This generalization can be readily adapted to the original MUSICAL indicator function given by Eq.~(\ref{eq:indicator-function}) by using the following assignments for $a_i$ and $b_i$, with $a_i + b_i=1$: 
\begin{equation} \label{eq:ai}
a_i = \begin{cases}
        1 & \text{if } \mathbf{u}_i \in \mathcal{S} \\ 
        0 & \text{otherwise}
      \end{cases}
\qquad
b_i = \begin{cases}
        1 & \text{if } \mathbf{u}_i \in \mathcal{N}\\ 
        0 & \text{otherwise}
      \end{cases}
\end{equation}

\subsection{Indicator function with eigenvalue (EV) weighing}
This indicator function follows $a_i + b_i=\lambda^{-1}$ and is defined as:
\begin{equation}
a_i = \begin{cases}
        {\lambda_i^{-1}} & \text{if } \mathbf{u}_i \in \mathcal{S} \\ 
        0 & \text{otherwise}
      \end{cases}
\qquad
b_i = \begin{cases}
        {\lambda_i^{-1}} & \text{if } \mathbf{u}_i \in \mathcal{N} \\ 
        0 & \text{otherwise}
      \end{cases}
\label{eq:ev}
\end{equation}
 This indicator function design retains both KC1 and KC2, but does not use the Euclidean projections $||{\mathbf{g}_\mathcal{S}(\mathbf{r}_{\rm test})}||$ and $||{\mathbf{g}_\mathcal{N}(\mathbf{r}_{\rm test})}||$ on the signal and noise subspaces. 
It instead weighs the projections on individual eigenimages $g_i$ according to the inverse of its singular value. This is graphically illustrated in Fig.~\ref{fig:threshold_maps}a for rule B. 

We present further insight into the EV indicator function using Fig.~\ref{fig:threshold_maps}c-d. 
Due to multiplication with the inverse of eigenvalues, $g_i$ for any order gets amplified for background regions and attenuated for the foreground regions. The foreground attenuation helps the higher order eigenimages to better support the resolution and the lower orders in reducing the dynamic range of the nanoscopy image on the higher side. The background amplification helps the higher order eigenimages to better suppress the background artifacts and the lower orders in reducing the dynamic range on the lower side. Thereon, the effect of hard thresholding is still present. Nonetheless, a significant softening is achieved as described next. Consider the orders that which are assigned to the signal subspace using Rule A but to the noise subspace using Rule B, however when treated using EV indicator function. When included in the signal subspace, they reduce the dynamic range of the original version of MUSICAL. This is significant because when using rule A, the original indicator function of MUSICAL generally supports better resolution but has extremely high indicator function values for few foreground pixels in nanoscopy image. This results into some pixels being highly saturated in the MUSICAL images and dynamic range of the image is not well-utilized, as reported in the supplement of the original MUSICAL paper \cite{agarwal2016multiple}.
On the other hand, when included in the noise subspace they help in improving the resolution, which may have been compromised in rule B as discussed before in section \ref{sec:thresh_ori}.

The proposed indicator function is inspired from the EV formulation reported previously for inverse source problems, for example in \cite{10.5555/1822462,10.5555/573302}. 
The similarity between these previously reported EV formulations and the one proposed here is limited to the denominator component of Eq.~(\ref{eq:indicator-function-general}) when combined with $b_i$ defined in Eq.~(\ref{eq:ev}). 
The use of the signal subspace in the numerator and the application of the indicator function on one sliding window at a time are unique to the MUSICAL architecture, first reported in \cite{agarwal2016multiple} while incorporation of EV weighing in $a_i$ in Eq.~(\ref{eq:ev}) is proposed for the first time here. 

\subsection{Indicator function with soft threshold (MUSICAL-S)}

An alternative approach is to use continuous functions for $a_i$ and $b_i$. 
Our proposed function is defined in Eq.~(\ref{eq:a-fuzzy}) and graphically illustrated in Fig. \ref{fig:threshold_maps}a. 
\begin{equation} \label{eq:a-fuzzy}
a_i(x) = \begin{cases}
            1 & \text{if $\sigma_i \geq \sigma_{\max}$}\\
            0 & \text{if $\sigma_i \leq \sigma_{\min}$}\\
            \dfrac{\log_{10}\sigma_i - \log_{10}\sigma_{\min}}{\log_{10}\sigma_{\max} - \log_{10}\sigma_{\min}} & \rm otherwise
        \end{cases}
\qquad; \qquad
b_i(x) = 1 - a_i(x)
\end{equation}
We enforce that $a_i + b_i=1$. In this equation $\sigma_{\max}$ and $\sigma_{\min}$ must be given. We pick those to be the maximum and minimum of the second singular values across all the windows in the image, since the change in the trend is strongly evident in the second singular values. We note that the choice also obviates the need of user specification of $\sigma_{\min}$ and $\sigma_{\max}$.
Thus, this approach includes both soft and automatic thresholding properties.

The design of indicator function as suggested above implies that the signal and noise subspaces are no longer orthogonal.
There are some eigenimages ($\mathbf{u}_i; \sigma_i\geq\sigma_{\max}$) that are relegated to the signal subspace $\mathcal S$ with full confidence ($a_i=1, b_i=0$). 
Similarly, there are some eigenimages ($\mathbf{u}_i; \sigma_i\leq\sigma_{\min}$) that are relegated to the noise subspace $\mathcal{N}$ with full confidence ($a_i=0, b_i=1$). 
For the remaining eigenimages, it is understood that these eigenimages may have non-negligible signal information even while being significantly affected by noise.
Therefore, these eigenimages are relegated to signal and noise space with reduced confidence ( indicated by non-extreme values of $a_i$ and $b_i$) while the behaviour and roles of the $g_i$ for lowest and the highest orders are unambiguous. 
The role of some intermediate orders such as order 8 is not explicit. When included in the signal space, it does not contribute resolution in foreground but may help in pushing the lower limit of the dynamic range up by enhancing the background. 
On the other hand, when included in the noise space, they may degrade the resolution but also pull the upper limit of the dynamic range down. By including them in reduced proportions in both denominators and numerators, we expect to strike a balance between the positive and negative aspects of their inclusion in the signal and noise subspaces and  thereby achieve a softening effect of the threshold.  

We note that this design is a significant deviation from the key concepts of MUSICAL. Since the signal and noise spaces now share some eigenimages, the KC1 defined in Section~\ref{sec:musical-theory} does not apply.
Furthermore, the KC2 has to be redefined as follows:

\vspace{1mm}
\noindent\textbf{Redefined KC2:} If the signal and noise subspaces are suitably defined,  then the $b_i$ weighted projection of $\mathbf{g}(\mathbf{r}_n) \forall n$ on the noise subspace $\mathcal{N}$ is small, which allows the denominator in Eq.~(\ref{eq:indicator-function-general}) to be small and the overall indicator function to be high at the emitter locations. 



\subsection{Indicator function for EV with soft-threshold (EV-S)}

The concept of soft-thresholding can be integrated in EV as well, as shown below: 
\begin{equation} \label{eq:a-fuzzy-ev}
a_i(x) = \begin{cases}
            {\lambda_i^{-1}} & \text{if $\sigma_i \geq \sigma_{max}$}\\
            0 & \text{if $\sigma_i \leq \sigma_{min}$}\\
            {\lambda_i^{-1}} \left(\dfrac{\log_{10}\sigma_i - \log_{10}\sigma_{min}}{\log_{10}\sigma_{max} - \log_{10}\sigma_{min}} \right) & \rm otherwise
        \end{cases}
\qquad ; \qquad
b_i(x) = \lambda_i^{-1} - a_i(x)
\end{equation}
As before, we enforce $a_i + b_i=\lambda^{-1}$. This design is graphically illustrated in  Fig.~\ref{fig:threshold_maps}a with the EV-S indicator function as the softest as compared to the other indicator functions, allowing smoother transition in $a_i$ and $b_i$. 
While EV alleviates the sensitivity to the threshold, MUSICAL-S removes the need for user-specified threshold even while reducing the sensitivity of noise on the overlapping region in Fig.~\ref{fig:threshold_maps}a. Since EV-S combines traits from both, it is expected to demonstrate reduced sensitivity and soft thresholding.

\subsection{Discussion of the proposed generalized framework}

Through the generalized indicator function, we have allowed for the creation of families of MUSICAL algorithms based on some design rules. 
Specifically, two families have been created. 

\noindent\textbf{Family based on coefficient relationship:} These are defined based on the relationship between the coefficients $a_i$ and $b_i$. The family $a_i + b_i = 1$ may be considered the original MUSICAL family while the family $a_i + b_i = \lambda_i^{-1}$ may be considered as EV family. Other families may also be designed similarly.

\noindent\textbf{Family of coefficient continuity:} These are defined based on the individual characteristics of $a_i$ and $b_i$. For example, in the family of hard threshold, the curve corresponding to $a_i$ experiences an abrupt jump at the threshold $\sigma_0$, and the signal and noise spaces are disjoint. In the family of soft threshold, the signal and noise subspaces are no longer orthogonal (KC1 does not apply) but the eigenimages in the overlapping space of signal and noise subspaces are weighted log-linearly. Other approaches may be designed for choosing the overlapping region or designing the weights, leading to other families of algorithms. 

Two final notes on the newly defined indicator functions:

\noindent\textbf{No resolution enhancement expected:} The aim of these new indicator functions is not resolution enhancement. In fact, no indicator function is expected to enhance or deteriorate resolution in particular. The only expected effect is a minor trade-off in resolution and contrast arising from different treatments of eigenimages in the signal and noise subspaces, as seen in Fig. \ref{fig:threshold_maps}e. 

\noindent\textbf{Removing subjectivity through automatic thresholding:} An important implication of the soft automatic indicator functions is that the subjectivity in threshold selection as well as the dependence on heuristics is removed. Outloook for further customization by an advanced user is possible, for example through a different choice of $\sigma_{\min}$ and $\sigma_{\max}$, or $a_i$ and $b_i$.

\section{Results}\label{sec:results}
We performed the following studies to compare the performances of the newly proposed indicator functions with the original indicator functions of MUSICAL: 

\begin{itemize}\itemsep0em 
    \item Quantitative analysis: We compare resolution and contrast for the indicator functions. We consider the effect of intensity fluctuations (determined by the photon emission on/off time) and the signal to background ratio (SBR). 2D samples are used so that other effects such as out-of-focus light do not affect the quantitative results.
    \item Qualitative analysis: Different 3D structures seen in biological samples are simulated and analyzed to study how the different indicator functions deal with realistic 3D structures and out-of-focus light. 
    \item Results on experimental data: We show comparative results using experimental data of a diverse set of samples. The dynamic range coverage is also investigated.
\end{itemize}

We consider a total of six indicator functions referred to as MUSICAL A, MUSICAL B, MUSICAL-S, EV A, EV B and EV-S. The value of $\alpha=4$ is used as recommended in \cite{agarwal2016multiple} and 10 subpixels per pixel are considered sufficient for the investigations presented in this article.

\vspace{1mm}\textbf{Simulation methods for studies 1-3:} 
The simulation involves the following steps: simulating the geometry of the structures, the emitters distribution over the structures, and the photokinetics of each emitter using the photoemission model of \cite{10.1371/journal.pone.0161602}, applying the Gibson-Lanni PSF \cite{Gibson:92, Li:17} of the microscope to simulate the raw noise-free image stacks, and then simulating the noise in raw image stack using the noise model in \cite{agarwal2016multiple}. 
We have simulated images for an epifluorescence microscope of numerical aperture 1.42, pixel size 80 nm, and emission wavelength of the emitters as 665 nm. 
Thereby, the theoretical resolution limit is approximately 285 nm using Rayleigh criterion \cite{novotny_hecht_2012}. For each sample, 500 frames were generated with 10 ms of exposure time for each frame. Duty cycle is considered as the ratio of the average time a fluorophore is emitting light ($\tau_{on}$) and the total cycle ($\tau_{on} + \tau_{off}$). Photobleaching is not considered. \donotshow{The samples and simulation code are available at \url{???}. }Where not mentioned explicitly, SBR of 4 and duty cycle of 5\% are utilized.

\subsection{Quantitative analysis}
\begin{figure}[t]
    \centering
    \includegraphics[width=\linewidth]{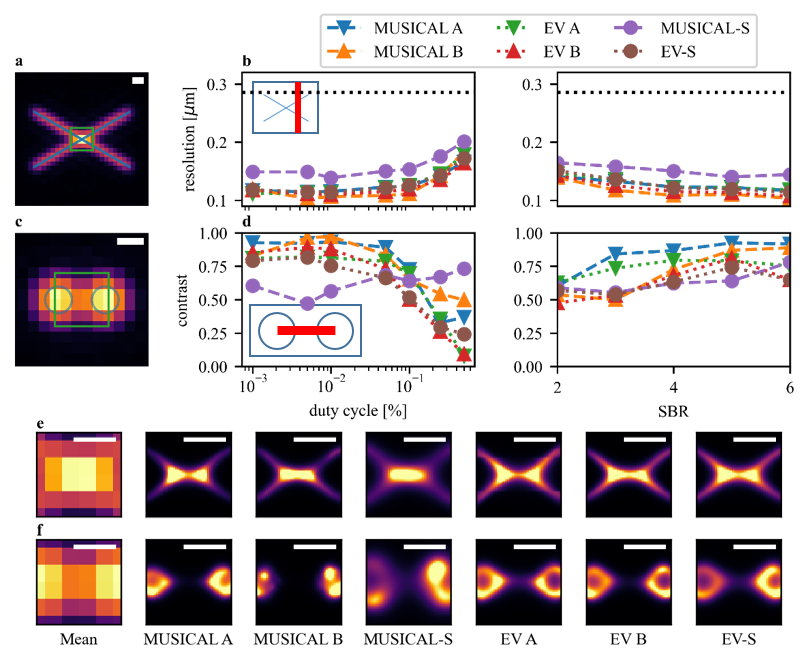}\vspace{-3mm}
    \caption{\textbf{Resolution and contrast study}. \textbf{a,c.} Studied samples with the groundtruth in blue and green square shown magnified in panels \textbf{e,f} for qualitative illustration. \textbf{b,d}, Resolution and contrast, respectively,  for different methods at different duty cycles and SBR. Each data point in Fig. \textbf{b,d} corresponds to the mean value over 100 independent simulations. Dashed line in \textbf{b} shows the resolution limit. Scale bar is 200 nm.}
    \label{fig:quantitative}
\end{figure}

\noindent\textbf{Study of resolution:}
This study is performed on a sample comprising two crossing lines with an angle of 60° as shown in Fig.~\ref{fig:quantitative}a, where the emitters are uniformly distributed at a density of  500 per $\upmu$m.
The results of Fig.~\ref{fig:quantitative}b show the distance from the crossing point at which the 2 lines can be differentiated according to the Rayleigh criterion \cite{novotny_hecht_2012} for different values of the duty cycle and SBR.
The diffraction limited resolution is shown as a blue line for reference.

The resolution is estimated by computing the ratio between valley and peaks across a moving vertical section (shown in the inset of Fig.~\ref{fig:quantitative}b), starting from the intersection point ($x = 0$). Let $\mathbf{l}(x)$ denote the image's intensities across a vertical line passing by $x$. We compute the ratio:
\begin{equation}
    r(x) = \frac{v(\mathbf{l}(x))}{min(p_1(\mathbf{l}(x)), p_2(\mathbf{l}(x)))}
    \label{eq:ratio}
\end{equation}
Here, $v$ and $p_i$ are functions that return the minimum and maximum intensity value respectively value around the expected valley and peaks $i$ (bottom and top peaks). 
The reported resolution is given as the minimum value $x$ at which $r(x)\leq 0.835$.
The range considered for duty cycle goes from 0.1 \% (comparable to single molecule localization microscopy data) up to 50\%. 
This last case corresponds to a highly dense spatio-temporal emission situation, which is a challenging situation for most fluctuations based techniques \cite{2008.09195}. Further, we consider SBR from 2 to 6, where SBR 2 is considered quite poor.

From Fig.~\ref{fig:quantitative}, the first observation is that all the methods provide resolution enhancement by a factor between 2 and 3. 
When comparing rules, we find that the best results are obtained using rule B for both MUSICAL and EV. 
Considering hard thresholding and comparing the MUSICAL and EV families, MUSICAL takes the lead in general.
Among soft-threshold methods, EV-S performs comparable to MUSICAL and EV while but MUSICAL-S displays the worst performance among all of them.
The poorer resolution of MUSICAL-S is also observed in Fig~\ref{fig:quantitative}c where the borders of the central cross appear diffused compared to the other methods, producing a lower resolution.
Lastly, we note that duty cycles below 10\% and SBR higher than 4 do not provide significant improvement in resolution. 

\vspace{1mm}
\noindent\textbf{Study of contrast:}
We here consider a sample containing emitters distributed uniformly across the perimeter of two circles of diameter 200 nm (density of 500 per $\upmu$m) at a distance of 150 nm between edges.
The mean image and groundtruth are shown in Fig~\ref{fig:quantitative}c. 
The contrast is defined as $c= 1 - r(\mathbf{l}(0))$, where $r(x)$ is defined in Eq.~(\ref{eq:ratio}) and $\mathbf{l}(0)$ represents the intensities across the horizontal section between circles (see inset of Fig. \ref{fig:quantitative}d). Higher values of $c$ indicate better contrast.
The results in Fig~\ref{fig:quantitative}d indicate that the contrast for all the indicator functions deteriorates as the duty cycle increases, with a larger slope after 10\%. 
The only exception is MUSICAL-S which shows low sensitivity to duty cycle and has relatively flat contrast for all duty cycles.
In terms of SBR, similar to the line sample, the curves become flat for SBR $\geq$ 4.
When comparing rules, rule A achieves marginally better contrast both in MUSICAL and EV, with MUSICAL performing the best.
The soft-threshold methods showed all similar and poorer results.

\vspace{1mm}
\noindent\textbf{Summary of the quantitative study:} The soft-thresholding schemes provide resolution and contrast similar to the other hard thresholding schemes. In general, EV-S outperformed MUSICAL-S in terms of resolution and contrast due to the weighting scheme by including singular values. Additionally, {from the results we observe that above a SBR of 4 and and duty cycle of 10\%, no improvement in resolution or contrast were achieved}.

\subsection{Qualitative Examples}
\label{sec:qualitative-examples}

\begin{figure}
    \centering
    \includegraphics[width=\linewidth]{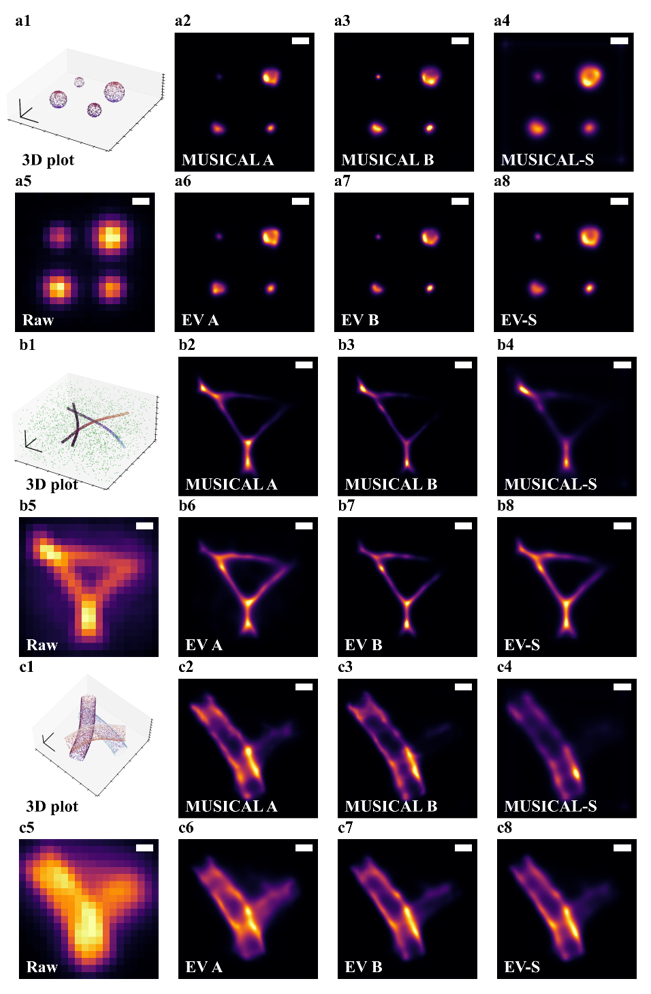}
    \caption{\textbf{Qualitative study on synthetic samples.} Three different samples are shown.  \textbf{a.} 4 vesicles of different size. \textbf{b.} Three crossing microtubules with debris. \textbf{c.} The crossing mithondria. All scale bars are 300 nm.}
    \label{fig:qualitative_large}
\end{figure}

\vspace{1mm}\noindent\textbf{Vesicles with surface labeling (Fig~\ref{fig:qualitative_large} a1-a8):} Four vesicles of different sizes are simulated as spheres of diameters 150, 200, 250 and 300 nm. Membrane labeling is simulated by distributing emitters on surface with density 800 per $\upmu$m$^2$. 
The vesicles are placed such that their centers are at the focal plane. Larger vesicles result with certain portions out of focus.

In MUSICAL A (Fig~\ref{fig:qualitative_large} a3), the smallest and therefore dimmer structure (top-left vesicle) is almost invisible. 
This is not the case for MUSICAL B (panels a4) which is explained
by a lower cardinality of the signal space and therefore better discarding of noise.
In the case of MUSICAL-S ( Fig~\ref{fig:qualitative_large} a5) the structure is even more visible, and the remaining vesicles looks more uniform.
No clear difference is observed across EV reconstructions (Fig~\ref{fig:qualitative_large} a6, a7 and a8).

It is of interest to observe that for the largest vesicle (300 nm in diameter), 
the out of focus light is rejected by all the methods, producing a hollow in the middle that is not visible for the microscopy image.
Below that size, the entire vesicle can be considered to be in focus.
\donotshow{We observe that even if the size of the circle reconstructed is proportional to the original radius, this does not match the correct size when using full width at half maximum (FWHM).}

\vspace{1mm}\noindent\textbf{Microtubules with background debris (Fig.~\ref{fig:qualitative_large} b1-b8):} Microtubules are fiber-like polymers of tubulin monomers. Fluorescent dyes label the monomers, which may be present as freely dispersing in addition to microtubule fiber \cite{nogales2001structural}.
This results in fluorescent debris, which is generally unwelcome in reconstruction. 
For this sample, we simulated fibers of 30 nm in diameter, with dyes distributed randomly across their surface at a linear density of 800 emitters per $\upmu$m.
Additionally, debris is added as single emitters distributed randomly with a volumetric density of 1000 emitters per $\upmu m^3$.
Both, microtubule-bound and free emitters, are assumed to have the same photo-kinetics.  
The geometry consists of three crossing lines forming an inverted triangle when seen from the top (Fig.~\ref{fig:qualitative_large} b2).
In the top-left and bottom corner, the structures meet in the focal plane, while the microtubules meeting at the top-right corner are both out of focus and separated by 500 nm in axial direction.
The spatial distribution can be seen in Fig.~\ref{fig:qualitative_large} b1.

\begin{figure}
    \centering
    \includegraphics{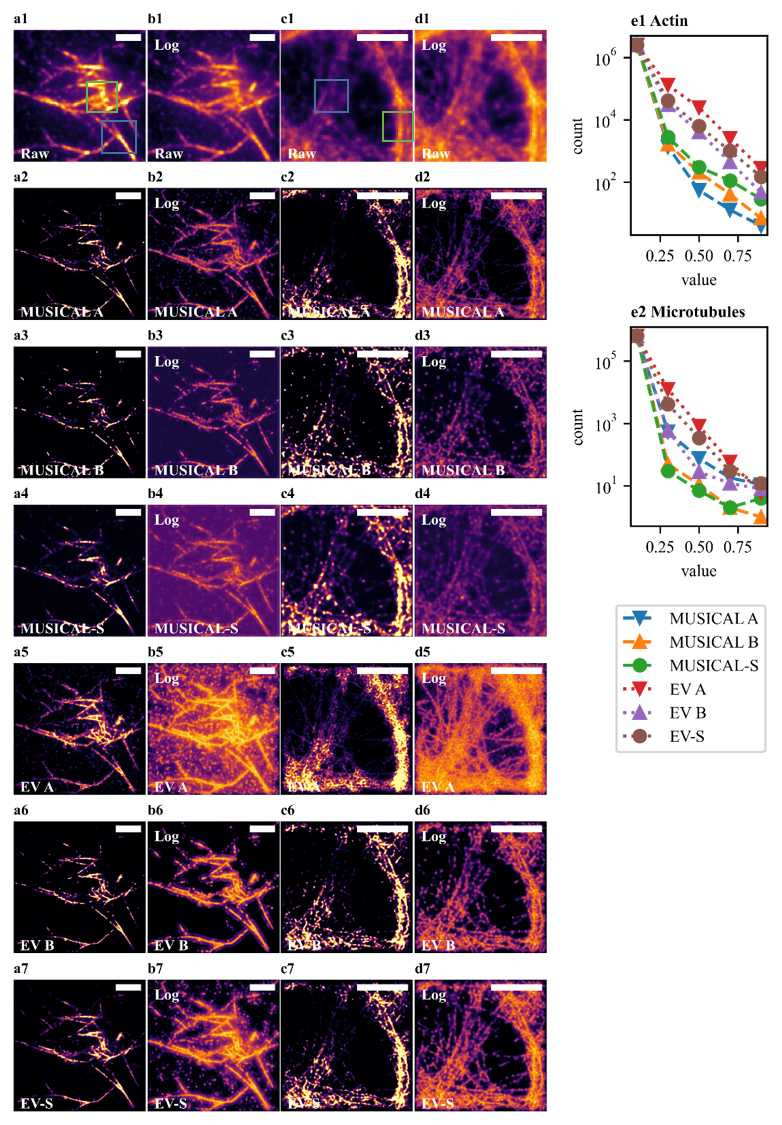}
    \caption{\textbf{Results on experimental data}. The results are displayed in two scales: linear (\textbf{a} and \textbf{c}) and logarithmic with base 10 (\textbf{b} and \textbf{d}). Blue and green boxes show regions of interest. The scale bar is 2 $\upmu$m in every image. \textbf{a},\textbf{b}. Results for actin (invitro). \textbf{c}, \textbf{d}. Results on microtubules. \textbf{e.} Comparative plot of the intensities obtained from the histogram of each normalized image by its maximum.}
    \label{fig:experimental}
\end{figure}

The results show that debris is absent in reconstructions regardless of the method while the overall structure is well reconstructed.  
A difference in the reconstruction of regions with out-of-focus structures is noticeable:
MUSICAL A , B and S (Fig~\ref{fig:qualitative_large} b3, b4 and b5 respectively) achieve a better rejection than their EV counterparts (Fig~\ref{fig:qualitative_large} b6, b7 and b8). 
However, the corners of the triangle which are the brightest points do not allow a clearer reconstruction of the strands due to the dynamic range problem of MUSICAL reported in \cite{agarwal2016multiple} and discussed in section \ref{sec:new-method}.
In this sense, a better reconstruction is obtained by EV. 
It shows that a desired level of rejection of out-of-focus structures could determine the choice between MUSICAL and EV. 

\vspace{1mm}\noindent\textbf{Mitochondrion (Fig.~\ref{fig:qualitative_large} c1-c8):} Mitochondria are tubular structures with diameters close to or larger than the diffraction limit.
Even if such structures are in focus, the sometimes large diameters combined with the dynamic nature of these organelles in living samples, causes large portions of the samples to be out of focus in realistic microscopy experiments.
Here, we consider an example of three mitochondria with diameter 300 nm and a density of emitters of 3000 per $\upmu$m$^2$ on the outer membrane. 
Each mitochondrion is in a different plane, with the left most mitochondrion in the focal plane. 
The top one is in the plane 300 nm above (closer to the coverslip) and the last one is 300 nm below the focal plane (further from the coverslip).
As in the case of microtubules, the out of focus rejection is  prominent for all methods with MUSICAL obtaining the strongest rejection of out of focus signal. Only the portions in the focal plane are reconstructed with good contrast. Further, the structure away from the coverslip is rejected less effectively than the structure above, which can be explained by the asymmetry of the PSF. 

\vspace{1mm}\noindent\textbf{Summary of this study:} The MUSICAL indicator functions were found to perform stronger out-of-focus rejection than the EV indicator functions. 

\subsection{Results on experimental data}

The methods were here tested on real microscopy data of filamentous actin and microtubules in fixed cells.
The samples used for testing the methods correspond to invitro actin filaments \cite{agarwal2016multiple} and microtubules in fixed cells \cite{Agarwal2018}. 500 frames were used for each reconstruction. 

\vspace{1mm}\noindent\textbf{Invitro actin filaments (Fig.~\ref{fig:experimental} a,b):} This sample is thin in the sense that that all the structures can be considered in focus. We marked two regions of interest (blue and green boxes) where super-resolution can give better insight of the structure than conventional microscopy. In the blue box (close to bottom), the bifurcation is clearly visible using the different indicator functions, and almost no difference is observed in terms of structure.
However, the region in the green box is reconstructed better in EV as they show a better distribution of the intensities. 
This allows to observe the entire network of strands without saturating the colors in other regions and is mainly attributed to the contrast enhancement due to the softening effect of EV. On the other hand in the images in logarithmic scale (Fig.~\ref{fig:experimental}b), we observe how EV introduces artifacts in the background, which is reduced when using a larger threshold such as the one used with rule B. 
In the case of the soft thresholding methods, EV-S looks crisper and more defined that MUSICAL-S, while providing better contrast between foreground and background.
The pixel distribution is plotted in log scale in Fig.~\ref{fig:experimental} a8, where it can be clearly observed how EV produces an intensity distribution that is higher in the middle tones, making better use of the dynamic range.

\vspace{1mm}\noindent\textbf{Microtubules in fixed cells (Fig~\ref{fig:experimental} c,d):}
This sample of fixed cell is different from the actin sample in the sense that some structures are expected to be out-of-focus. For images in linear scale (Fig.~\ref{fig:experimental} c1-c7), the methods display similar  performance, except for EV A. 
In particular, the blue region (left box) presents a high degree of artifacts, where it is difficult to visualize individual strands due to high saturation. The same occurs in the green region, where EV A recovers just one single structure, while all the remaining methods manage to recover 2 strands.This illustrates an example of the minor trade-off between resolution and contrast. The better contrast and visibility, as well as poorer rejection of out-of-focus structures is strongly evident in the log scale (Fig. \ref{fig:experimental}d). 
Between MUSICAL-S and EV-S, the latter produces a better result by achieving better contrast and definition.
Both methods recover the dim structures in the right-most box.

\section{Conclusion}\label{sec:conclusion}

Through a generalized framework for MUSICAL indicator function design, we have proposed new indicator function families and specific indicator function designs to address problem of hard threshold in MUSICAL. 
The EV family of indicator functions soften the effect of threshold and provide better utilization of the dynamic range of MUSICAL.
This is generally achieved at no cost of resolution but poorer rejection of out-of-focus light as compared to the MUSICAL family of indicator functions.
Further, a soft threshold family is proposed that does not define signal and noise subspaces of MUSICAL as strictly orthogonal, and allows an overlap between them. 
Therefore, it removes the concept of hard thresholding, however, altering the key concepts of MUSICAL. 
While MUSICAL-S indicate poorer resolution, the EV-S indicator function which combines the traits of both EV and soft-threshold families shows consistently good results across a wide variety of quantitative, qualitative, and experimental studies. 
Through this work, we widen the horizon for MUSICAL in two important aspects.
First, the sensitivity and subjective choice of threshold is removed which makes it easier to use. Second, it opens exciting avenues for further development of fluctuations based super-resolution algorithms in general, and MUSICAL in particular. 
In the future, it will be interesting to design customized indicator functions for challenging scenarios such as dynamic samples where the different methods can be tested.


\bibliography{references}

\end{document}